\input{aipcheck}
\def\ltap{\ \raise.3ex\hbox{$<$\kern-.75em\lower1ex\hbox{$\sim$}}\ }
\newcommand{\ket}[1]{|{#1}\rangle}
\newcommand{\bra}[1]{\langle{#1}|}

\documentclass[
    ,final            
  ]
  {aipproc}

\layoutstyle{8x11single}

\begin{document}

\title{Neutrino-induced meson productions off nucleon at forward limit
in nucleon resonance region}



\author{S. X. Nakamura}{
  address={Yukawa Institute for Theoretical Physics, Kyoto University, Kyoto 606-8542, Japan}
}

\author{H. Kamano}{
  address={Research Center for Nuclear Physics, Osaka University, Ibaraki, Osaka 567-0047, Japan}
}

\author{T.-S. H. Lee}{
  address={Physics Division, Argonne National Laboratory, Argonne, Illinois 60439, USA}
}

\author{T. Sato}{
  address={Department of Physics, Osaka University, Toyonaka, Osaka 560-0043, Japan}
  ,altaddress={J-PARC Branch, KEK Theory Center, Institute of Particle and Nuclear Studies, High Energy Accelerator Research Organization (KEK), Tokai, Ibaraki 319-1106, Japan}
}

\begin{abstract}
We study forward neutrino-induced meson production off the nucleon in the
 resonance region.
Our calculation is based on a dynamical coupled-channels (DCC) model that
 reasonably describes
$\pi(\gamma) N\to \pi N, \eta N, K\Lambda, K\Sigma$ data in the
 resonance region. 
We apply the PCAC hypothesis to the DCC model to relate the $\pi N$
 reaction amplitude to the forward neutrino reaction amplitude.
In this way, we give a prediction for 
$\nu N\to \pi N, \pi\pi N, \eta N, K\Lambda, K\Sigma$ reaction cross
sections.
The predicted
$\nu N\to \pi\pi N, \eta N, K\Lambda, K\Sigma$ cross sections are,
 for the first time, based on a model extensively tested by data. 
We compare our results with those from the Rein-Sehgal model that has
 been very often used in the existing Monte Carlo simulators for
 neutrino experiments. 
We find a significant difference between them.
\end{abstract}

\maketitle


\section{Introduction}

A groundbreaking measurement of non-zero $\theta_{13}$ opened a new
avenue for neutrino oscillation experiments. 
Next generation experiments will be targeting the leptonic CP violation
and the mass hierarchy of the neutrino.
To achieve this goal, it is essential to understand the neutrino-nucleus
interaction more precisely, 10\% or even better,
over a rather wide kinematical region that covers quasi-elastic, resonance,
and deeply inelastic scattering (DIS) regions.

In this contribution, we are concerned with the resonance region, from
the $\Delta(1232)$ through 2nd and 3rd resonance regions, up to $W\ltap 2$~GeV.
Several models have been developed for 
the neutrino-induced single pion production off the nucleon in the
resonance region, and have been used as basic ingredients to construct
neutrino-nucleus interaction models.
Some of them considered coherent sum of resonance contributions~\cite{rs1,rs1-2,lalakulich1,lalakulich1-2}.
Some others additionally took account of non-resonant mechanism of the
tree level~\cite{hernandez1,hernandez1-2,lalakulich2}.
Two of the present authors further considered the rescattering also so
that the $\pi N$ unitarity is maintained~\cite{sl3,msl05}.
So far, most models deal with only the single pion production.
However, the neutrino-nucleon interaction in the resonance region is a
multi-channel reaction.
Two-pion production has a comparable contribution to the single pion
production.
$\eta$ and kaon productions can also happen. 
In order to deal with this kind of multi-channel reaction, 
an ideal approach is to develop a unitary coupled-channels model; this
is what we will pursue.

Recently we have been developing a unitary dynamical
coupled-channels (DCC) model that can be extended to the neutrino reactions~\cite{aip11,knls12}.
Our DCC model is based on a comprehensive analysis of 
$\pi N,\gamma N\to \pi N, \eta N, K\Lambda, K\Sigma$ reactions in the
resonance region, taking account of the coupled-channels unitarity including 
the $\pi\pi N$ channel.
In this contribution, we report our first step of extending the DCC
model to the neutrino reaction~\cite{pcac}. 
For that, we invoke the Partially Conserved Axial Current (PCAC) hypothesis that allows us to relate 
cross sections of the pion-induced meson productions  to those of the
corresponding neutrino-induced meson productions at forward limit. 

In what follows, we give a brief description of our DCC model, 
and formulate the neutrino reaction cross section
with the PCAC hypothesis.
Then we show our result together with a comparison with the Rein-Sehgal
model.

\section{Dynamical coupled-channels model}

\begin{figure}
  \includegraphics[width=10cm]{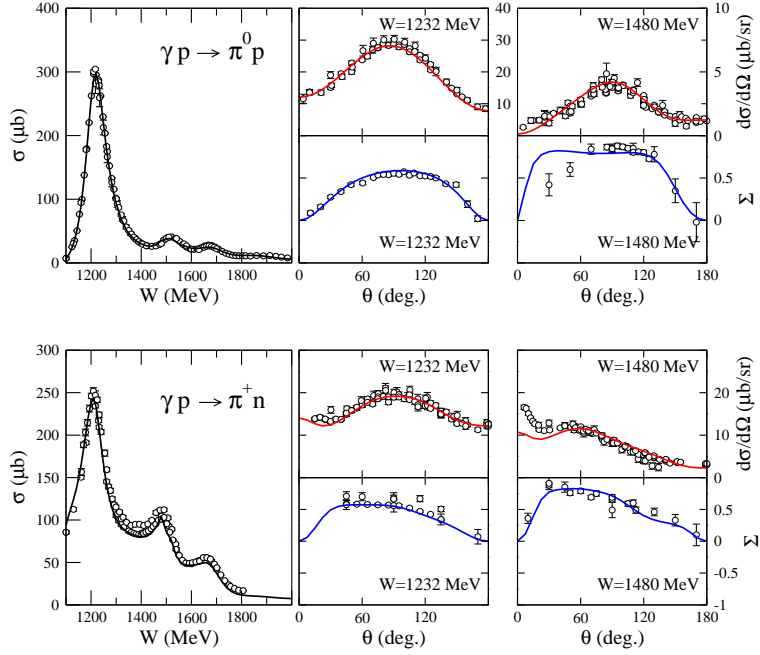}
  \caption{\label{fig1} Comparison of the DCC model~\cite{aip11,knls12}
 and data for single pion photoproduction.
The upper [lower] figures are total cross sections ($\sigma$), unpolarized
 differential cross sections ($d\sigma/d\Omega$) and photon asymmetry
 ($\Sigma$) for $\gamma p \to \pi^0 p$ [$\gamma p \to \pi^+ n$].
The total energy is denoted by $W$, and the scattering angle of the
 pion by $\theta$.
}
\end{figure}

In our DCC model~\cite{aip11,knls12,msl}, 
we consider 8 channels: 
$\gamma N, \pi N, \eta N, \pi\Delta, \rho N, \sigma N, K\Lambda, K\Sigma$.
The $\pi\pi N$ channel is included in the 
$\pi\Delta, \rho N, \sigma N$ channels using Feshbach's projection
method, thus maintaining the three-body unitarity.
Meson-exchange driving terms are derived from 
meson-baryon Lagrangian.
The driving terms as well as bare $N^*$ excitation mechanisms are
implemented in a coupled-channel Lippmann-Schwinger equation from which 
we obtain unitary reaction amplitudes.
We analyzed 
$\pi(\gamma) N\to \pi N, \eta N, K\Lambda, K\Sigma$ reactions data 
simultaneously up to $W = 2.1$~GeV ($W$ : total energy).
The analysis includes fitting about 20,000 data points.
As an example for showing the quality of the fit,
we show in Fig.~\ref{fig1}
the single pion photoproduction observables from the DCC model
compared with data.
As shown in the figure, 
our DCC model gives a reasonable description of meson production
data in the resonance region.
As a consequence, the DCC model contains all four-star resonances (and
more) listed by the Particle Data Group~\cite{pdg}.
Thus the DCC model provides a good basis with which we proceed to the neutrino reactions.

\section{PCAC and neutrino-induced forward meson productions}

Kinematic variables used in the following discussion are as follows.
We consider the inclusive $l (k) + N(p) \to l'(k') + X(p')$ reactions
($X=\pi N, \pi\pi N, \eta N$... etc.), where 
$(l,l') = (\nu_e, e^-), (\bar \nu_e, e^+)$ for the charged-current (CC)
reactions.
Although we do not show a result, the neutral-current reaction can be
studied in a similar manner.
We assume that leptons are massless.
In the laboratory frame, the four-momentum are defined to be
$k=(E, \vec k)$, $p=(m_N, 0, 0, 0)$, $k'=(E', \vec k')$
and $p' = k + p -k'$.
With the momentum transfer between $l$ and $l'$, $q= k'-k =(\omega,\vec q)$, 
we define the positive quantity $Q^2$ by
$Q^2 = -q^2 = 4 E E^\prime \sin^2\frac{\theta}{2}$,
where $\theta$ is the scattering angle of $l'$ with respect to $l$
 in the laboratory frame.

For later use, we also define another frame where $X$ is at rest.
In this frame, $q$ and $p$ are denoted as $ q=(\omega_c,\vec q_c)$ and 
$p = (E_N, -\vec q_c)$, respectively, where $E_N = \sqrt{m_N^2 + |\vec q_c|^2}$ 
and $m_N$ is the nucleon mass. Also, we set $\vec q_c =(0,0,|\vec q_c|)$ so that
$\vec q_c$ defines the $z$-direction of this frame.

The cross sections for the inclusive neutrino and anti-neutrino reactions are expressed as
\begin{equation}
\frac{d\sigma_\alpha}{dE^\prime d\Omega^\prime}
=
\frac{G^2_F V^2_{ud}}{2\pi^2}E^{\prime 2}
\left[ 
2W_{1,\alpha} \sin^2\frac{\theta}{2}
+W_{2,\alpha} \cos^2\frac{\theta}{2}
\pm W_{3,\alpha} \frac{E+E^\prime}{m_N}\sin^2\frac{\theta}{2}
\right] \ ,
\label{eq:crs1}
\end{equation}
where the label $\alpha = {\rm CC}\nu,{\rm CC}\bar\nu$
specifies the reactions;
$\Omega'$ is the solid angle of $l'$ in the laboratory frame;
$V_{ud}$ is the CKM matrix element;
the sign in front of $W_{3,\alpha}$ is taken to be $+$ ($-$) for $\nu$ ($\bar\nu$)
induced reactions.
The structure functions, $W_{i,\alpha}$ ($i=1,2,3$), are Lorentz-invariant functions of 
two independent variables, $(Q^2, W)$, where $W$ is the total energy of $X$
at its rest frame.
In the forward limit, $\theta \to 0$, Eq.~(\ref{eq:crs1}) reduces to
\begin{equation}
\frac{d\sigma_\alpha}{dE^\prime d\Omega^\prime} (\theta \to 0)
=
\frac{G^2_F V^2_{ud}}{2\pi^2}E^{\prime 2} W_{2,\alpha}  \,.
\label{eq:crs2}
\end{equation}

The structure function $W_{2,\alpha}$ is expressed in terms of matrix elements of 
weak currents between the initial nucleon $N$ and the final state $X$,
$\bra{X}J^\mu_\alpha\ket{N}$, as
\begin{equation}
W_{2,\alpha} 
=
\frac{Q^2}{\vec q^2} \sum 
\left[ \frac{1}{2}\left( |\bra{X} J_\alpha^x \ket{N}|^2 + |\bra{X} J_\alpha^y \ket{N}|^2 \right)
+ \frac{Q^2}{\vec q_c^2} \left|\bra{X}
\left(J_\alpha^0 + \frac{\omega_c}{Q^2} q\cdot J_\alpha \right)
\ket{N}\right|^2 \right] \,,
\label{eq:w2-1}
\end{equation}
where the summation symbol indicates all possible final states $X$,
integration over momentum states of $X$, the average of initial
nucleon spin state, and some kinematical factors including the
phase-space factor. 
In the forward limit where $Q^2=0$, 
what survives in Eq.~\ref{eq:w2-1} is only the last term that contains the
divergence of the current.
The weak current consists of the vector ($V^\mu$) and axial ($A^\mu$) currents.
Because of the vector current conservation $\bra{X} q\cdot V\ket{N}=0$ 
in the isospin limit, and $|\vec q_c| = \omega_c$ at $Q^2 = 0$. Thus we find
\begin{equation}
W_{2,\alpha}(Q^2 \to 0) = \frac{1}{\vec q^2} \sum  |\bra{X} q\cdot A \ket{N}|^2\,.
\label{eq:w2-2}
\end{equation}
According to Refs.~\cite{mf1,mf2,mf3},
we can define the pion field with
the divergence of the axial currents as
\begin{equation}
\bra{X(p')} q \cdot A^a \ket{N(p)}= 
f_\pi m_\pi^2 \bra{X(p')} \hat \pi^a \ket{N(p)} ,
\label{eq:pcac1}
\end{equation}
where $f_\pi$ ($m_\pi$) is the pion decay constant (pion mass), and 
$\hat\pi^a$ is the normalized interpolating pion field with the isospin
state $a$.
Furthermore, the matrix element $\bra{X(p')}\hat\pi^a\ket{N(p)}$ at $Q^2=0$ can be expressed as 
\begin{equation}
\bra{X(p')} \hat \pi^a \ket{N(p)} = 
\frac{\sqrt{2\omega_c}}{m_\pi^2} {\cal T}_{\pi^a N \to X} (0) .
\label{eq:pcac2}
\end{equation}
Here, ${\cal T}_{\pi^a N \to X}(q^2)$ is the T-matrix element
of the $\pi^a(q) + N(p) \to X(p')$ reaction in the $\pi N$ center-of-mass frame,
and the incoming pion is off-mass-shell $q^2 = 0 \not= m_\pi^2$.
Using Eqs.~(\ref{eq:pcac1}) and~(\ref{eq:pcac2}), we have at $Q^2=0$,
\begin{eqnarray}
\frac{1}{\vec q^2} \sum
|\bra{X(p')} q \cdot A^a \ket{N(p)}|^2 
 &=& \frac{1}{\vec q^2} f_\pi^2 (2\omega_c)\sum  |{\cal T}_{\pi^a N \to X}(0)|^2
\nonumber\\
&\sim&
 \frac{1}{\vec q^2} f_\pi^2 (2\omega_c)\sum  |{\cal T}_{\pi^a N \to X}(m_\pi^2)|^2 
\nonumber\\
&=&
 \frac{1}{\vec q^2} f_\pi^2 (2\omega_c) \frac{1}{2\pi}\frac{p\cdot q}{E_N\omega_c} \frac{E_N}{m_N}
\sigma_{\pi^a N \to X}
\nonumber\\
&=&
 \frac{f_\pi^2}{\pi\omega} \sigma_{\pi^a N \to X},
\end{eqnarray}
where $\sigma_{\pi^a N \to X}$ is the total cross section of 
the on-shell $\pi^a + N \to X$ reactions, 
and we have used the relation
 ${\cal T}_{\pi^a N \to X}(q^2 = 0)\sim {\cal T}_{\pi^a N \to X}(q^2=m_\pi^2)$,
which is a consequence from the PCAC hypothesis
and $\vec q^2 = \omega^2$.
Converting the isospin basis to the charge state basis,
we finally have
\begin{equation}
W_{2,\alpha} = 
\left\{
\begin{array}{cl}
\displaystyle \frac{2f_\pi^2}{\pi\omega }\sigma_{\pi^{+} N \to X}  
&({\rm for }\ \ \alpha ={\rm CC}\nu), \\
\\
\displaystyle \frac{2f_\pi^2}{\pi\omega }\sigma_{\pi^{-} N \to X}  
&({\rm for }\ \ \alpha ={\rm CC}\bar\nu). \\
\end{array}
\right.
\label{eq:w2-pcac}
\end{equation}
In the next section, we show the dimensionless structure function $F_2$
defined by $F_2=\omega W_2$.
The above result is essentially the same as those given
by Adler~\cite{adler} and also by Paschos {\it et al.}~\cite{pas1,pas2,pas3}.
From Eqs.~(\ref{eq:crs2}) and~(\ref{eq:w2-pcac}), one can evaluate 
neutrino-induced forward meson production cross sections at $\theta = 0$
using the $\pi N \to X$ total cross sections.

\section{Result}

\begin{figure}
  \includegraphics[height=.25\textheight]{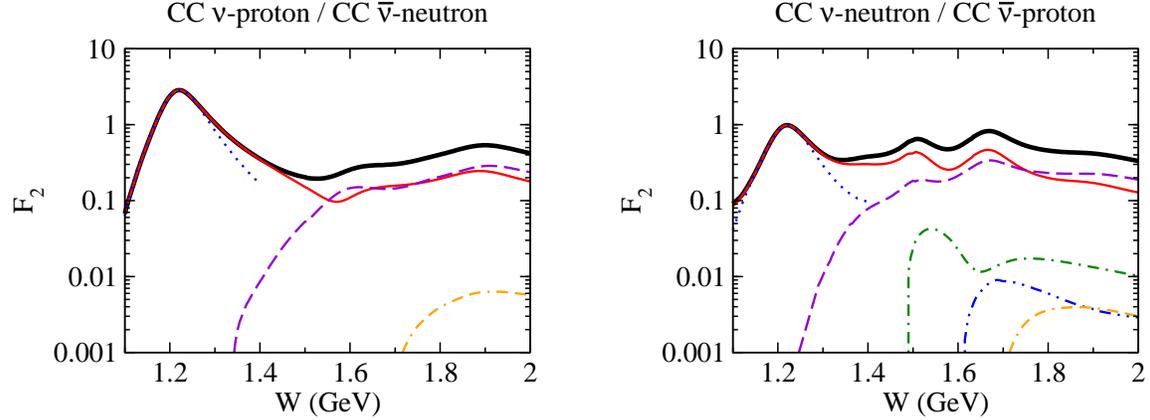}
  \caption{\label{fig2} The structure function $F_2 (Q^2=0)$ for the
 neutrino-induced meson productions from the DCC model. 
The solid (red), dashed (purple), dash-dotted (green), two-dotted
 dash (blue), and two-dash dotted (orange) curves are for the $\pi N$,
$\pi\pi N$, $\eta N$, $K\Lambda$ and $K\Sigma$ reactions, respectively.
The sum of them is given by the thick solid (black) curve. The SL model
is shown by the dotted (blue) curve. The figures are from Ref.~\cite{pcac}.}
\end{figure}

We show the structure functions $F_2$ for the neutrino-induced meson
productions off the nucleon in Fig.~\ref{fig2}.
The left panel of Fig.~\ref{fig2} shows CC neutrino-proton or
antineutrino-neutron scattering where only $I=3/2$ states ($I$ : total
isospin of final state $X$) contribute,
while the right panel is for CC neutrino-neutron or
antineutrino-proton scattering for which both $I=1/2$ and $3/2$ states
give contributions.
While $\pi N$ production is the dominant process up to 
$W=1.5$ GeV, 
above that energy, 
the $\pi \pi N$ production becomes comparable to $\pi N$, showing the
importance of the $\pi\pi N$ channel in the resonance region above $\Delta(1232)$.
Also, we observe that the $\pi N$ and $\pi\pi N$ spectra above the
$\Delta$ have rather bumpy structure, reflecting contributions from many
nucleon resonances. 
This structure cannot be simulated by a naive extrapolation of the DIS
model to the resonance region, as has been often done in previous
analyses of neutrino oscillation experiments.
Other meson productions, $\eta N$, $K\Lambda$, and $K\Sigma$ reactions
have much smaller contribution, about
[$O(10^{-1})$-$O(10^{-2})$] of $\pi N$ and $\pi \pi N$ contributions.

It is interesting to compare the above result for $X=\pi N$ with 
the counterpart from the Sato-Lee (SL) model~\cite{sl3,msl05}, as done in Fig.~\ref{fig2}.
The SL model directly gives the $F_2$ functions (without the PCAC hypothesis) because it
consists of both the vector and axial currents, and
reasonably reproduce available neutrino-induced pion production data in
the $\Delta(1232)$ region~\cite{sl3}.
From the comparison, we can see that contributions of nucleon
resonances above $\Delta$ is clearly important above $W = 1.3$ GeV.
Also, the good agreement 
$W\ltap 1.3$ GeV indicates
a reliability of calculating $F_2$ 
from the $\pi N \to X$ total cross sections with the PCAC hypothesis.
We remark that this is the first prediction of 
the neutrino-induced $\pi\pi N$, $\eta N$, $KY$ production rates based
on a model that has been extensively tested by data. 

\begin{figure}
  \includegraphics[height=.25\textheight]{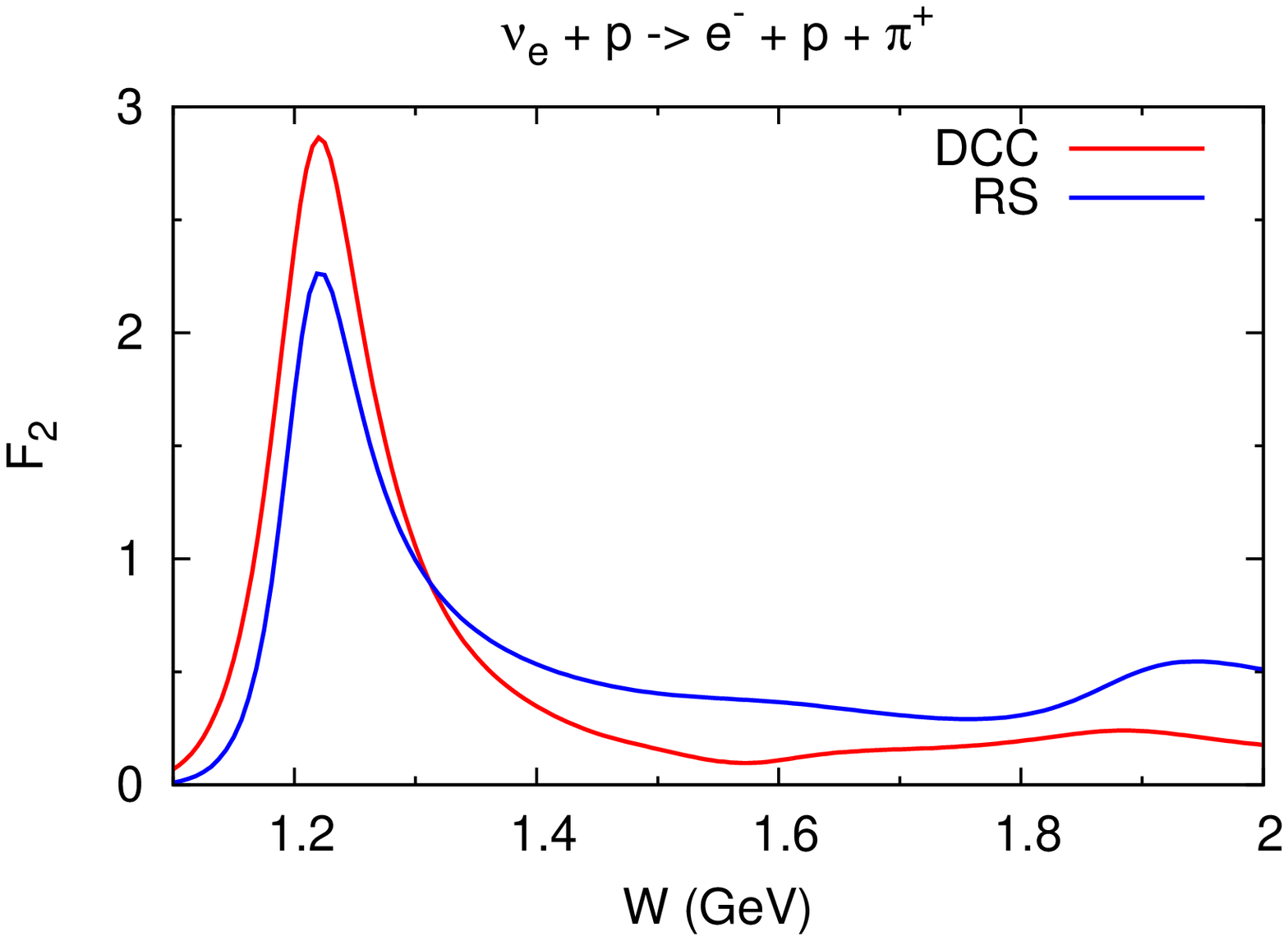}
  \includegraphics[height=.25\textheight]{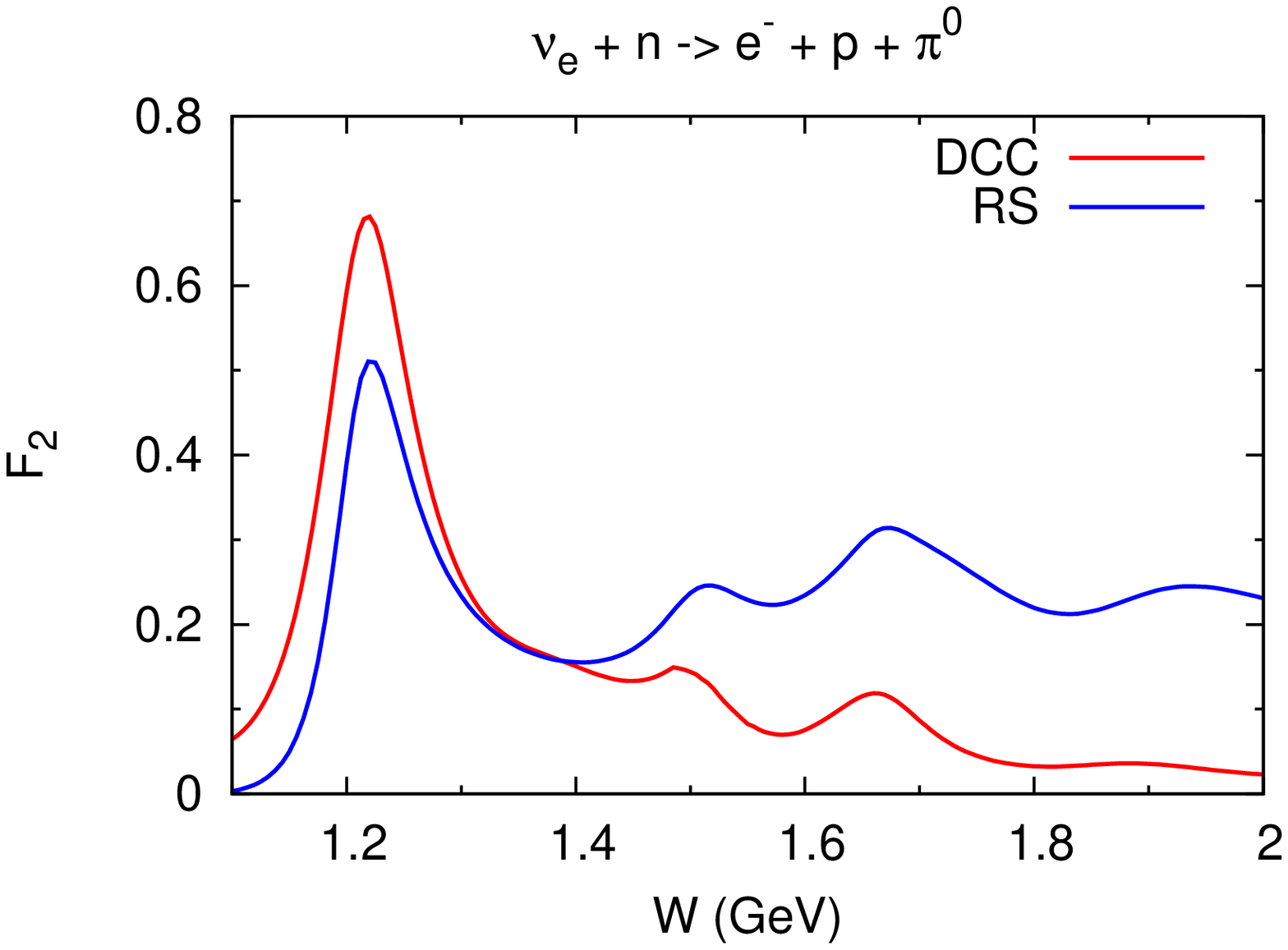}
  \caption{\label{fig3} Comparison of $F_2 (Q^2=0)$ between the DCC
 model and Rein-Sehgal (RS) model.}
\end{figure}

It is also interesting to compare our result with $F_2$ from the Rein-Sehgal
(RS) model~\cite{rs1,rs1-2} that has been used in many Monte Carlo simulators for
analyzing neutrino experiments. 
Such a comparison is shown in Fig.~\ref{fig3}.
We can see that the RS model underestimates the $\Delta(1232)$ peak by
$\sim$ 20\%. 
On the other hand,
in higher energies, the RS model significantly overestimates $F_2$.
Our result is based on the DCC model tested by lots of data in the
resonance region while the RS model has not but based on a quark model.
Considering that,
the current Monte Carlo simulators using the RS model should be improved. 
In this work, the comparison with the RS model is done only at the
forward limit. 
More comparison for non-forward kinematics, as well as full description
of neutrino reaction needs development of a dynamical axial current
model.
Such a development is currently underway.



\begin{theacknowledgments}
SXN is the Yukawa Fellow and his work is supported in part by Yukawa Memorial Foundation,
the Yukawa International Program for Quark-hadron Sciences (YIPQS),
and by Grants-in-Aid for the global COE program 
``The Next Generation of Physics, Spun from Universality and Emergence'' from MEXT.
HK acknowledges the support by the HPCI Strategic Program 
(Field 5 ``The Origin of Matter and the Universe'') of 
Ministry of Education, Culture, Sports, Science and Technology (MEXT) of Japan.
TS is supported by JSPS KAKENHI (Grant Number 24540273).
This work is also supported by the U.S. Department of Energy, Office of Nuclear Physics Division, 
under Contract No. DE-AC02-06CH11357.
\end{theacknowledgments}



\bibliographystyle{aipproc}   




\end{document}